# Collective skyrmion motion under the influence of an additional interfacial spin-transfer torque

Callum R. MacKinnon[1,*], Katharina Zeissler[2,3], Simone Finizio[4], Jörg Raabe[4], Christopher H. Marrows[2,3], Tim Mercer[1], Philip R. Bissell[1] & Serban Lepadatu[1,**]

[1] *Jeremiah Horrocks Institute for Mathematics, Physics and Astronomy, University of Central Lancashire, Preston PR1 2HE, U.K.*

[2] *School of Physics and Astronomy, University of Leeds, Leeds LS2 9JT, UK*

[3] *Bragg Center for Materials Research, University of Leeds, Leeds LS2 9JT, UK*

[4] *Swiss Light Source, Paul Scherrer Institut, 5232 Villigen, Switzerland*

**Abstract**

Here we study the effect of an additional interfacial spin-transfer torque, as well as the well-established spin-orbit torque and bulk spin-transfer torque, on skyrmion collections – group of skyrmions dense enough that they are not isolated from one another – in ultrathin heavy metal / ferromagnetic multilayers, by comparing modelling with experimental results. Using a skyrmion collection with a range of skyrmion diameters and landscape disorder, we study the dependence of the skyrmion Hall angle on diameter and velocity, as well as the velocity as a function of diameter. We show that inclusion of the interfacial spin-transfer torque results in reduced skyrmion Hall angles, with values close to experimental results. We also show that for skyrmion collections the velocity is approximately independent of diameter, in marked contrast to the motion of isolated skyrmions, as the group of skyrmions move together at an average group velocity. Moreover, the calculated skyrmion velocities are comparable to those obtained in experiments when the interfacial spin-transfer torque is included. Our results thus show the significance of the interfacial spin-transfer torque in ultrathin magnetic multilayers, which helps to explain the low skyrmion Hall angles and velocities observed in experiment. We conclude that the interfacial spin-transfer torque should be considered in numerical modelling for reproduction of experimental results.

[*]CRMackinnon@uclan.ac.uk, [**]SLepadatu@uclan.ac.uk

# Introduction

Recently, significant efforts have been dedicated to studying how magnetic skyrmions, topologically protected magnetic textures [1], can be stabilised at room temperature using the Dzyaloshinsky-Moriya interaction (DMI) [2,3], and displaced in ultrathin magnetic multilayers with electrical currents [4-10]. Even though skyrmions have clear advantages for future spintronic devices [11-13] over domain walls, in terms of information carrying and storage, there are still inherent challenges that skyrmions face in the transition to applications. The skyrmion Hall effect (SkHE) is one such challenge, and whilst it may play an important role in skyrmion logic gates [12,14], owing to the transverse motion that it gives rise to, it also causes the skyrmion to traverse to the edges of a track leading to possible annihilation [15]. Whilst the SkHE may be reduced, or even eliminated, in synthetic ferrimagnetic and antiferromagnetic structures [16,17], there remains a fundamental interest in understanding the quantitative dependence of the SkHE on skyrmion diameter, driving forces, and disorder.

Driving skyrmions in ultrathin magnetic multilayers via charge current injection utilises the spin-Hall effect (SHE) [8,18], which generates a pure spin current from an unpolarized charge current. The spin-orbit coupling in the heavy metal (HM) layer gives rise to an asymmetric scattering of conduction electrons, and thus an asymmetric deflection of spin-up and spin-down electrons in opposite directions, creating a transverse spin current. Opposing spins accumulate at surfaces of the HM layer, giving rise to a spin-orbit torque (SOT) on the adjacent ferromagnetic (FM) layer [19-23], which causes rotation of the local spin directions and thus skyrmion motion. In most works the SOT is the principal mechanism considered for displacing skyrmions, however it has been shown that another torque arising in the FM layer, from the spin accumulated at magnetization gradients [24,25], such as a skyrmion or domain wall, can also have an important effect on skyrmion motion, namely the interfacial spin-transfer torque (ISTT). The ISTT arises due to an imbalance in spin accumulation at the interface between HM and FM layers. This imbalance leads to diffusive vertical spin currents away from the FM layer, such that the transverse spin components are absorbed at the interface, resulting in a spin torque due to conservation of total angular momentum. Whereas the SOT arises from absorption of the transverse spin components carried by a vertical spin current originating in the HM layer, the ISTT arises from a vertical spin current originating in the FM layer. The ISTT has been shown to have a significant effect on skyrmion motion in terms of the SkHE and velocity [25], and could explain the observation of a skyrmion Hall angle (SkHA) independent of the skyrmion diameter [18].

It is well known that the movement of isolated skyrmions, as well as skyrmion collections (group of skyrmions dense enough to undergo collective dynamics), is strongly influenced by disorder, pinning sites, and boundaries [26]. Skyrmions moving near the boundary of a magnetic multilayer have also been shown to be repelled away from the boundary [27]. Furthermore, skyrmions interacting with neighbouring skyrmions have also been shown to repel each other due to the dipole-dipole and exchange interactions, thus leading to the possibility of skyrmion-skyrmion interaction [28,29], and constraint of skyrmion diameter when a layer is saturated by skyrmions [30]; the extreme case being a skyrmion crystal. Moreover, the interaction of a skyrmion collection with the boundaries of a multilayered track was recently shown to result in a reshaped SkHE [31]. Driving individual skyrmions through a disordered landscape has been extensively studied under the effects of SOT from a HM layer [5,32], spin transfer torque (STT) in the FM layer [33], and more recently ISTT [24,25].

In this work we study the effect of the additional ISTT term, alongside the STT and SOT, on skyrmion collections – defined as a group of interacting skyrmions – in HM/FM multilayers with landscape disorder in the form of surface roughness. Modelling results are compared to experimental result on SkHA and velocities, for skyrmion collections with a range of diameters. We show through micromagnetic simulations coupled to a self-consistent drift-diffusion spin transport solver, that the SkHA-diameter dependence is in quantitative agreement with experimental results, contrary to modelling results which consider the SOT as the sole driving spin torque, particularly for small skyrmion diameters. Moreover, we also show skyrmion collections move approximately at an average group velocity, independent of diameter, and the experimental velocities are close to those obtained from modelling when the ISTT is included, whilst SOT-only modelling again shows discrepancies with experimental data.

## Spin transport model and interfacial spin torques

The effect of spin torques on collective skyrmion motion is modelled using the Landau-Lifshitz-Gilbert (LLG) equation, in which the effective field term consists of several contributions, namely the demagnetizing field, exchange interaction, interfacial DMI, applied magnetic field, and uniaxial magnetocrystalline anisotropy. Furthermore, we can introduce additive spin torque terms to the LLG, which include contributions from SOT, ISTT, and STT, and which can be computed self-consistently using the drift-diffusion model, as shown previously in more detail [24,25,34].

Within the drift-diffusion model, interfacial spin torques may be computed using the spin-mixing conductance, $G^{\uparrow\downarrow}$, from the spin accumulation **S** either side of the interface:

$$\mathbf{T}_S = \frac{g\mu_B}{ed_F}\left[\mathrm{Re}\{G^{\uparrow\downarrow}\}\mathbf{m}\times(\mathbf{m}\times\Delta\mathbf{V}_S) + \mathrm{Im}\{G^{\uparrow\downarrow}\}\mathbf{m}\times\Delta\mathbf{V}_S\right]. \tag{1}$$

Here $d_F$ is the ferromagnetic layer thickness, $\Delta\mathbf{V}_S$ is the spin chemical potential drop across the HM/FM interface, where $\mathbf{V}_S = (D_e/\sigma)(e/\mu_B)\mathbf{S}$, with $D_e$ the electron diffusion constant and $\sigma$ the electrical conductivity. To determine the contribution due to the SOT, the drift-diffusion model can be solved analytically to obtain the following expression, by assuming negligible in-plane spin diffusion:

$$\mathbf{T}_{SOT} = \theta_{SHAeff}\frac{\mu_B}{e}\frac{|J_C|}{d_F}\left[\mathbf{m}\times(\mathbf{m}\times\mathbf{p}) + r_G\mathbf{m}\times\mathbf{p}\right]. \tag{2}$$

Here $\mu_B$ is the Bohr magneton, $e$ the electron charge, $\mathbf{p} = \mathbf{z}\times\mathbf{e}_{Jc}$, where $\mathbf{z}$ is a unit vector perpendicular to the film plane, $\mathbf{e}_{Jc}$ is the charge current density direction, with $|J_C|$ being its magnitude, and **m** is the magnetization unit vector. The SOT term contains both damping-like (DL) and field-like (FL) components, where $r_G$ is the FL-SOT coefficient. The term $\theta_{SHAeff}$ is the effective spin-Hall angle which is proportional to the intrinsic spin-Hall angle. The distinction between the effective and intrinsic spin-Hall angle at the HM/FM interface is important, as shown previously [35], as the former can be significantly smaller depending on the interface.

The ISTT arises due to HM/FM interlayer diffusion of a spin accumulation generated in the FM layer, which creates another important source of vertical spin current. The ISTT contribution has a similar form to the bulk Zhang-Li STT [36,37], but is significantly larger due to the inverse FM thickness dependence in Equation (1), given by ($M_S$ is the saturation magnetization):

$$\mathbf{T}_{ISTT}/M_S = -[(\mathbf{u}_{ISTT}.\nabla)\mathbf{m} - \beta_{ISTT}\mathbf{m}\times(\mathbf{u}_{ISTT}.\nabla)\mathbf{m}]. \tag{3}$$

The spin-drift velocity, $\mathbf{u}_{ISTT} = |P_{ISTT}|g\mu_B\mathbf{J}_{FM}/2eM_S(1+\beta_{ISTT}^2)$, and non-adiabaticity parameter $\beta_{ISTT}$ are effective interfacial terms that can differ significantly from their bulk counterparts. The inverse FM thickness dependence results in large adiabatic and non-adiabatic spin torques. For the work presented here we have $P_{ISTT}$ = -1.34 and $\beta_{ISTT}$ = -0.96, where the effective ISTT polarization is $\tilde{P}_{ISTT} = P_{ISTT}/(1+\beta_{ISTT}^2) = -0.69$, obtained by fitting the ISTT expression in Equation (3) to the self-consistent spin torque, Equation (1), computed using the drift-diffusion model – for details see Ref. [25] and the Supplementary Information. The bulk STT values are $P$ = 0.42 and $\beta$ = 0.002.

In this work we also consider the effect of the bulk STT. Within the drift-diffusion model, bulk spin torques are obtained from the spin accumulation as:

$$\mathbf{T}_S = -\frac{D_e}{\lambda_J^2}\mathbf{m}\times\mathbf{S} - \frac{D_e}{\lambda_\varphi^2}\mathbf{m}\times(\mathbf{m}\times\mathbf{S}). \tag{4}$$

Here $\lambda_J$ is the exchange rotation length, and $\lambda_\varphi$ is the spin dephasing length. This bulk spin torque reduces to the well-known bulk STT, with spin-drift velocity $\mathbf{u}$ and non-adiabaticity parameter $\beta$ [34]. The SOT, STT, and ISTT are incorporated into the LLG equation to give a fuller description of the magnetization dynamics for magnetic textures in multilayers as:

$$\begin{aligned}\frac{\partial\mathbf{m}}{\partial t} &= -\gamma\mathbf{m}\times\mathbf{H}_{eff} + \alpha\mathbf{m}\times\frac{\partial\mathbf{m}}{\partial t}\\ &+ \theta_{SHAeff}\frac{\mu_B}{e}\frac{|J_c|}{M_S d_F}[\mathbf{m}\times(\mathbf{m}\times\mathbf{p}) + r_G\mathbf{m}\times\mathbf{p}]\\ &+ [(\mathbf{u}-\mathbf{u}_{ISTT}).\nabla]\mathbf{m} - \mathbf{m}\times[(\beta\mathbf{u}-\beta_{ISTT}\mathbf{u}_{ISTT}).\nabla]\mathbf{m}\end{aligned} \tag{5}$$

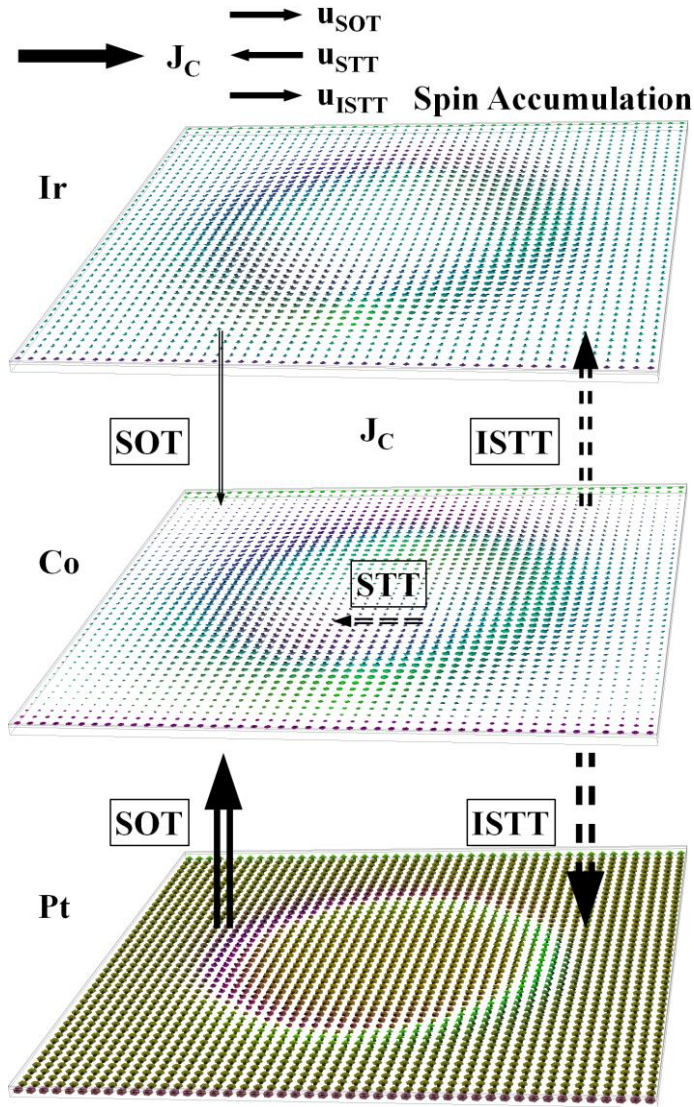

**Figure 1**. Diagrammatic representation of the origins of different spin torques in a Pt/Co/Ir multilayer. The SHE gives rises to vertical spin currents (indicated by solid arrows) flowing from Pt (and to a much smaller extent Ir) into Co, resulting in SOT. Spin accumulation at the skyrmion in Co gives rise to diffusive vertical spin currents (dashed arrows), flowing from Co into Pt and Ir, resulting in ISTT. SOT and ISTT are obtained from the self-consistent interfacial spin torque – Equation (1). Spin accumulation at the skyrmion in Co also gives rise to in-plane spin currents (small dashed arrow), resulting in the much weaker bulk STT – computed self-consistently using Equation (4).

Fig. 1 shows a diagrammatic representation of the different spin torques arising in a Pt(2.7 nm)/Co(0.8 nm)/Ir(0.4 nm) stack, with a charge current density in the plane as indicated. The computed spin accumulation in the three layers is shown. In Pt and Ir the SHE contributes to the generated spin accumulation; vertical spin currents flowing from Pt to Co (and to a much smaller extent from Ir to Co), give rise to SOT. On the other hand, a spin accumulation is generated in Co due to magnetization gradients at the skyrmion. This gives rise to a total STT with bulk and interfacial components. In-plane spin currents give rise to the Zhang-Li bulk STT component, however this is much smaller than the SOT and ISTT – a discussion of the different spin torques is given in the Supplementary Information. Diffusive vertical spin currents flowing from Co into Pt and Ir, give rise to ISTT. It must be emphasized, the ISTT (just like bulk STT) requires magnetization gradients to be present, thus for example in ferromagnetic resonance (FMR) experiments on multilayers, where the magnetization is approximately uniform, ISTT is negligible.

Experimental results are often compared to the analytical model known as the Thiele equation [38], which is a rigid body skyrmion approximation used to describe the dynamics of the skyrmion due to driving forces. For a skyrmion, the Thiele equation contains three contributions, namely, the force acting on the skyrmion, the SkHE, and the dissipative force. An expression for the diameter-dependent and drive-independent SkHA, $\theta_{SkHA}$, as well as the velocity, $v$, can be derived from the Thiele equation, with results given by:

$$v = \sqrt{\frac{(u_{SOT} - \beta_{STT} D u_{STT})^2 + u_{STT}^2}{1 + (\alpha D)^2}},$$

$$\tan(\theta_{SkHA}) = \frac{u_{SOT} + (\alpha - \beta_{STT}) D u_{STT}}{\alpha D u_{SOT} - (1 + \alpha \beta_{STT} D^2) u_{STT}}.$$
(6)

Here $u_{SOT} = \mu_B \theta_{SHAeff} \pi R J_{HM} / 4 e M_S d_F$, and $u_{STT} = P_{STT} g \mu_B J_{FM} / 2 e M_S (1 + \beta_{STT}^2)$, where $J_{HM}$ and $J_{FM}$ are the current densities in the HM and FM layers respectively, and $D = R/2\Delta$, where $\Delta$ is the domain wall width and $R$ is the skyrmion radius. Equation (6) models the contributions of the SOT, as well as the total STT (including bulk and interfacial contributions), and for current-induced skyrmion motion agrees closely with micromagnetic modelling [25]. The combination of bulk STT and ISTT in Equation (5) results in a total STT with parameters $P_{STT}$ and $\beta_{STT}$, obtained from bulk and interfacial components as $\beta_{STT} = (\beta \tilde{P} + \beta_{ISTT} \tilde{P}_{ISTT}) / (\tilde{P} + \tilde{P}_{ISTT})$, and $P_{STT} = (1 + \beta_{STT}^2)(\tilde{P} + \tilde{P}_{ISTT})$, where $\tilde{P}_{(ISTT)} = P_{(ISTT)} / (1 + \beta_{(ISTT)}^2)$.

For the purpose of studying skyrmion collections and their current-induced motion, we use Boris Computational Spintronics [39], a numerical micromagnetics package coupled with a self-consistent spin transport solver. We model a Pt(2.7 nm)/Co-like(0.8 nm)/Ir(0.4 nm) stack, with in-plane area of 2×2 µm$^2$ and periodic boundary conditions along the *x*-axis direction only, thus effectively simulating a wire oriented along the *x*-axis. Computations are done using cell-centered finite difference discretization. For magnetization dynamics, the computational cell size is (2, 2, 0.8 nm) – the exchange length is ~4 nm – and for spin transport calculations the cell size is (2, 2, 0.1 nm). The LLG equation is evaluated using the Runge-Kutta-Fehlberg (RKF45) method, whilst the spin accumulation is computed using the iterative successive over-relaxation method. Parameters for the Co-like layer are, exchange stiffness $A$ = 10 pJ/m, perpendicular anisotropy $K$ = 380 kJ/m, and saturation magnetisation $M_S$ = 600 kA/m.

A skyrmion tracking algorithm is used with multiple variable window sizes, which allows tracking any number of skyrmions. From the extracted skyrmion paths, the average SkHA and velocities may then be obtained for each skyrmion, as well as average values for the entire collection, which contains a range of skyrmion diameters.

Multilayer growth and wire fabrication

D.C. magnetron sputtering at a base pressure of 2×10$^{-8}$ mbar and a target-substrate separation of ~7 cm was used to deposit the thin film heterostructure with typical growth rates of around 0.1 nm/s. During the growth the argon pressure was 3.2 mbar. The heterostructure studied consists of Ta(3.2)/Pt(2.7)/[Co$_{68}$B$_{32}$(0.8)/Ir(0.4)/Pt(0.6)]$_{×8}$/Pt(2.2), (thicknesses are in nm). The patterned samples (2-µm-wide wires), fabricated using electron beam lithography, were grown on x-ray transparent, highly resistive 200 nm thick silicon nitride membranes (Silson Ltd, Warwickshire, UK). An identical thin film was simultaneously sputtered onto a thermally oxidized Si substrate with an 100 nm thick oxide layer to enable the characterisation of the materials properties using standard techniques. X-ray reflectivity was used to measure the layer thicknesses.

The 2-µm-wide wires were fabricated using a bilayer electron-beam-sensitive resist process with a bottom layer of methyl-methacrylate (MMA) and a top layer of polymethyl-methacrylate (PMMA). The spun and baked resist bilayer was exposed using a 100 kV Vistec EBPG 5000Plus electron beam writer with a writing dose of 1650 µC cm$^{-2}$. The pattern was developed for 90 s in a 1:3 methyl-isobutyl-ketone and isopropyl alcohol solution (by volume) and rinsed for 60 s in isopropyl alcohol. The unpatterned regions were lifted off in acetone following the heterostructure growth procedure described above. Thermally evaporated 200-nm-thick Cu electrodes were fabricated using lift off electron beam lithography with the same exposure procedure as for the magnetic wires. The electrodes were designed to achieve an electrical impedance close to 50 Ω, minimizing unwanted reflections of the injected current pulses.

The magnetic configuration of the wires was imaged using scanning transmission X-ray microscopy (STXM) at the PolLux beamline of the Swiss Light Source [40]. The X-ray magnetic circular dichroism (XMCD) effect [41] was utilized as contrast mechanism. The X-ray incident angle was chosen to be perpendicular to the sample, resulting in out of plane magnetization sensitivity. Skyrmions in the wire were nucleated after initial saturation at 80 kA/m, by 2 current pulses with a width of 22 ns and a peak current density of $4\times10^{11}$ A/m$^2$ at -8 kA/m, using a combination of thermal, spin torques, and field effects to randomly nucleate skyrmions in the nanostructure [42,43]. After the nucleation pulse the initial magnetic state was imaged. Then, the motion image sequence was taken by applying three 22-ns-long current pulses, separated by a delay of 2 μs, with a maximum current density of $4.6\times10^{11}$ A/m$^2$ and average value of $2.0\times10^{11}$ A/m$^2$. Subsequently, static single helicity STXM images of the magnetization state of the wire were taken. This sequence was repeated 19 times. The TrackMate algorithm [44] was used to identify the skyrmions' center coordinates and the best fit diameter. Using the change in the centre position of each skyrmion, its velocity and SkHA were evaluated. An example of this imaging procedure, and extracted skyrmion paths, is shown in the Supplementary Information. Brillouin light scattering (BLS) measurements, performed on samples grown in identical conditions and with the same layer thicknesses (but with 5 stack repetitions), have obtained a DMI constant of $D$ = -1.1 mJ/m$^2$ [18]. This exceeds the critical value to enforce homochiral Néel walls around spin textures. Moreover, the generated structures move with a finite SkHA when driven. This shows that they cannot be trivial bubbles, but must possess topological charge and hence a gyrovector. We also performed micromagnetic simulations using the same stack with 8 repetitions as used in experiments, and have confirmed the stabilization of homochiral Néel skyrmions in the multi-layered stack as well – details are given in the Supplementary Information

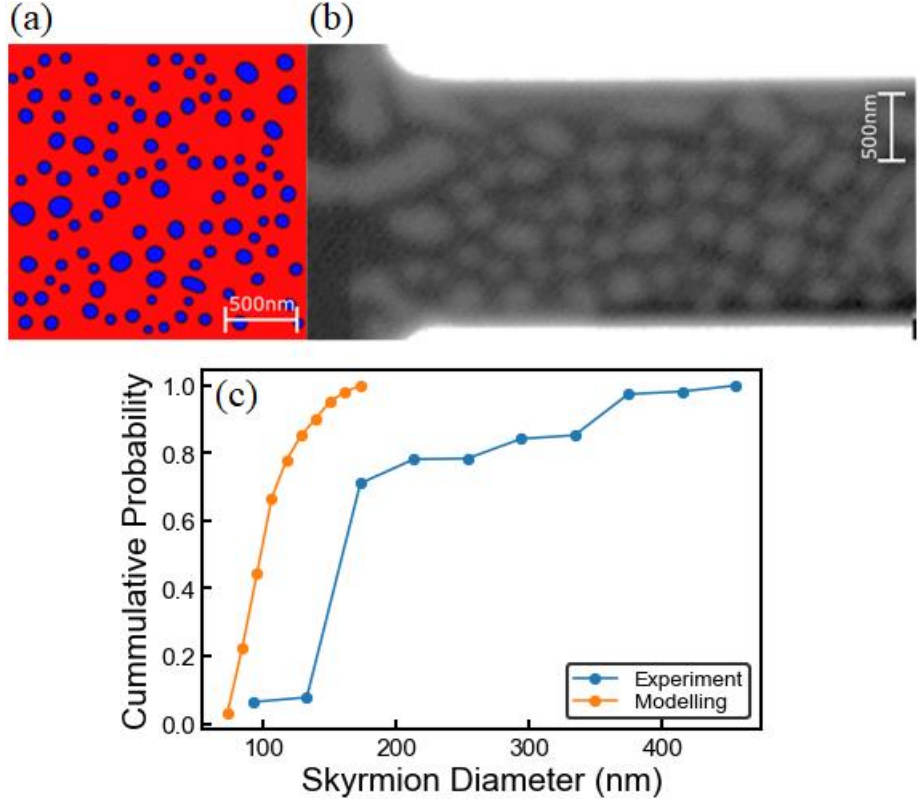

**Figure 2**. Skyrmion collections. (a) A simulated system with skyrmion density $N = 12.5$ Sk/µm$^2$, with a range of diameters stabilised by an out-of-plane field of $H = 1.2$ kA/m. Blue indicates in-plane magnetization and red out-of-plane. (b) STXM image of a skyrmion collection after a current pulse, $J_C = 4.5 \times 10^{11}$ A/m$^2$, in a 2 µm wide wire. Bright contrast indicates skyrmions. Images are shown at zero field. (c) Cumulative probability as a function of diameter, where around 75% of skyrmions in experiments are in the diameter range 70 – 180 nm.

Results

We first discuss the geometrical constraints on skyrmions in a collection, whilst relaxing the system to equilibrium state. A selected number of skyrmions were placed randomly within the FM mesh and relaxed using the steepest descent method, as shown in Fig. 2(a) for a skyrmion density of $N = 12.5$ Sk/µm$^2$. Periodic boundary conditions along the $x$-axis are incorporated into our model. This allows us to simulate over a much larger effective length, which is useful in the case of many skyrmions, but also allows for $y$-boundary interactions as in the experimental samples. Therefore, as the skyrmions are not interacting with the $x$-boundaries of the FM layer, the skyrmions can traverse large distances and can interact with the $y$-boundary. The skyrmion diameters are strongly influenced by adjacent skyrmions, in addition to $y$-boundary interactions, such that there is confinement of diameter as the sample becomes saturated with skyrmions [45]. The simulation space also incorporates landscape disorder, in the form of surface roughness, with 1 Å maximum depth and a 40 nm in-plane coherence length, as detailed previously [25]. During the relaxation phase, skyrmions collectively adjust to the most energetically favourable diameter, which is strongly influenced not only by interactions with neighbours, but crucially by landscape disorder which results in a variation of the effective out-of-plane anisotropy. The skyrmion-skyrmion interaction has been studied in great detail, and is known to result in a repulsive force between neighbouring skyrmions [46-48], dependent on the distance between them, the exchange interaction, and the DMI [28]. Without disorder the skyrmions relax to a regular hexagonal lattice. This is not the case when disorder is introduced, and a skyrmion collection will typically contain a range of diameters as indicated in Fig. 2(a),(b) – for example in Fig. 2(a) the diameters range from 70 nm to 180 nm, with a median of 110 nm. Here, the smallest diameter skyrmions tend to arrange into local hexagonal lattice clusters, with the larger diameter skyrmions showing significant distortions. This is in agreement with experimental images of skyrmion collections in the studied samples – for example Fig. 2(b) shows a STXM image of a skyrmion collection in a 2 μm wide wire. Here the diameters range from 72 nm to over 475 nm, with a median of 164 nm, and as for the simulated skyrmion collection the smaller skyrmions arrange into approximately hexagonal clusters, whilst the largest skyrmions appear significantly distorted and elongated.

We note that in current-driven simulations the average deviation from the mean skyrmion diameters is ~10 nm, due to interactions with the local landscape disorder. The skyrmion diameter during relaxation, as well as under an applied current, can experience distortions such that skyrmions no longer appear entirely circular. Two orthogonal profiles of the out-of-plane magnetization profiles are obtained through the skyrmion, which are fitted using $m_Z(x) = \cos(2\tan^{-1}(\sinh(R/\Delta)/\sinh((x-x_0)/\Delta)))$, where $R$, $\Delta$ and $x_0$ (center position) are fitting parameters. This procedure obtains the skyrmion diameters along two orthogonal in-plane directions, and the overall diameter is obtained as the average value. If the skyrmion is heavily distorted (the two orthogonal values differ by more than 50%), we no longer deem this a viable skyrmion and disregard it from calculations. For current-driven experiments, in total over 4500 skyrmions have been analysed, and ~75% of these are in the same diameter range as that obtained from modelling, as shown in Fig. 2(c).

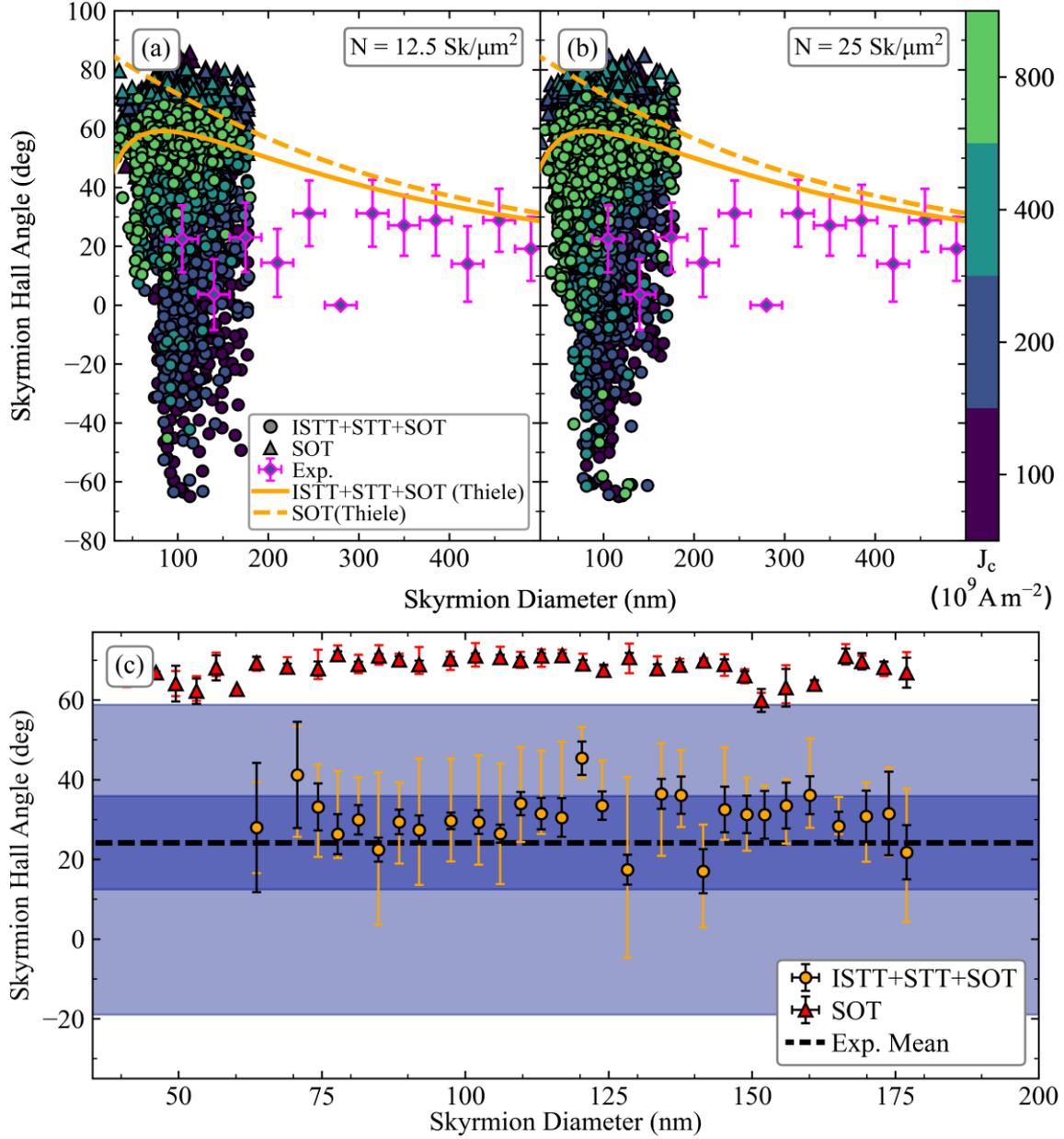

**Figure 3**. SkHA-diameter dependence for skyrmion collections with (a) $N = 12.5$ Sk/µm$^2$, and (b) $N = 25$ Sk/µm$^2$. The solid disks and triangles represent results for ISTT+STT+SOT, and SOT-only respectively, for a corresponding applied current density indicated by the colour bar. Magenta outlined diamonds represent the experimental results. The orange solid and dashed lines represent the Thiele equation solution for spin torque contributions ISTT+STT+SOT, and SOT-only respectively, with a damping value $\alpha = 0.1$. (c) Detail at the average current density used in experiments, $J_C = 2\times10^{11}$ A/m$^2$. The dashed horizontal line is the mean experimental SkHA, with the inner blue horizontal band showing standard error on the mean, and lighter blue outer band showing the spread (quartile 1 to quartile 3). This is compared to the solid disks showing modelling results with ISTT+STT+SOT, and solid triangles for SOT-only, also at $J_C = 2\times10^{11}$ A/m$^2$ (smaller error bars are standard error on the mean, and larger error bars are the spread).

There are three skyrmion motion regimes for a single skyrmion: (i) the pinned regime, in which skyrmions are pinned by the local energy landscape and cannot move, (ii) the depinning regime, in which skyrmions jump intermittently between pinning sites with a net motion, and (iii) the flow regime in which the motion is not heavily impeded by the local energy landscape and appears near linear [9]. The presence of these regimes is also found for a skyrmion collection, which we discuss in relation to Fig. 3. It should be noted that skyrmions in the pinned regime are not included, for all figures here and onwards, as they have a net zero velocity. For a skyrmion collection, the depinning regime can be subdivided into elastic and plastic depinning [49], the former occurring for weak disorder and being characterized by skyrmions retaining their original neighbours throughout the motion.

Modelling results shown in Fig. 3 are given for a skyrmion density of (a) $N = 12.5$ Sk/µm$^2$, and (b) $N = 25$ Sk/µm$^2$. The skyrmion density in experiments is estimated to be in the range 10 – 15 Sk/µm$^2$. At low current densities ($1.0\times10^{11} \leq J_C \leq 4.0\times10^{11}$ A/m$^2$), skyrmions are largely in the elastic depinning regime, and are strongly influenced by the local landscape disorder, as well as repulsive interactions with neighbouring skyrmions, resulting in a pronounced dependence of the SkHA on driving current density. This is especially evident for results obtained using the full spin torque (ISTT+STT+SOT), and to a lesser extent for the results obtained using SOT alone. For the full spin torque, due to the combination of ISTT, STT, and SOT, the net effect is a significantly smaller driving force for the same current density when compared to SOT-only, which helps to explain the more pronounced variation in SkHA under the effect of landscape disorder. The spread of SkHA is further enhanced for the skyrmion collection with higher density, observed by comparison of Fig. 3(a),(b), expected due to the increased effect of repulsive interactions with neighbouring skyrmions. Contributions of the individual spin torques are analysed in the Supplementary Information. At larger current densities ($J_C > 4.0\times10^{11}$ A/m$^2$) skyrmions are largely in the flow regime, with SkHA approximately described by the ideal SkHA-diameter dependence obtained using the Thiele equation. This is shown in Fig. 3 for the full spin torque as a solid line, and for the SOT-only model as a dashed line. The experimental SkHA values, bin-averaged over skyrmions in the collection for the observed diameter range, are shown in Fig. 3 as diamonds. At the current density used in experiments (average $J_C = 2.0\times10^{11}$ A/m$^2$), the skyrmions also move largely in the elastic depinning regime. Thus a large discrepancy arises between the results obtained using the Thiele model, although it should be noted this is more pronounced for the SOT-only model. This discrepancy persists for the SOT-only model even at large damping values of $\alpha = 0.3$, as discussed in the Supplementary Information, where the effect of varying the damping constant is also considered. Instead, to compare with experimental results, the SkHA values of skyrmions in the collection are shown in Fig. 3(c) for $J_C = 2.0 \times 10^{11}$ A/m$^2$, thus in the elastic depinning regime. The experimental SkHA is largely independent of skyrmion diameter, and the overall mean is indicated as a dashed line in Fig. 3(c). The inner blue band shows the standard error on the mean, whilst the outer lighter blue band shows the overall spread of experimentally measured SkHA values (defined as quartile 1 to quartile 3). Over 4500 skyrmions have been analysed in experiments, of which over 3500 are not pinned, and 75% of skyrmions are in the diameter range 70 – 180 nm, overlapping with the range obtained in modelling. A very good overlap is seen with the modelling results for the full spin torque, whilst for the SOT-only results a large discrepancy of over 50° is observed. Thus, whilst both the full spin torque, and SOT-only models reproduce the nearly diameter-independent SkHA trend qualitatively, only the full spin torque model is able to reproduce the SkHA values quantitatively. Moreover, the full spin torque model also reproduces the relatively large spread of experimental SkHA values in the skyrmion collection, as shown in Fig. 3(c), with an excellent overlap between the two regions. Also, the negative SkHA values observed in the spread of experimental results are only reproduced when the ISTT is included, which is due to the significant reduction in average SkHA, as well as reduction in velocities in the depinning regime. The spread of SkHA values is larger for skyrmion collections, compared to isolated skyrmion motion [25], since skyrmions interact not only with the local landscape disorder, but also with neighbouring skyrmions. As noted above, the spread in SkHA is reduced for the SOT-only model, since skyrmion velocities are significantly larger at the same current density (skyrmion velocities and comparison with experimental results are discussed later in relation to Fig. 5). The measured samples consist of 8 stack repetitions, thus 8 FM layers, whereas the simulated geometry contains 1 FM layer. A stack repetition has the effect of further reducing the SkHA, as discussed previously [24], both for SOT and ISTT+STT+SOT models, however this is not enough to explain the experimental SkHA values with the SOT-only model [24]. Indeed, as the stack repetition is increased, the skyrmion diameter increases due to the reduction in effective anisotropy [4]. However, this effect is much weaker for a skyrmion collection, where the skyrmion diameter is constrained by repulsive interactions with neighbouring skyrmions. We have calculated the average skyrmion diameter for a collection as a function of number of stack repetitions, and for 8 repetitions as used in experiment, the increase in the average skyrmion diameter is ~50 nm – this is shown in the Supplementary Information. As also shown there, a stack repetition does not significantly alter the spin torques, as the SOT is largely generated through the thicker Pt underlayer, and the ISTT is due to vertical spin currents flowing away from the Co layer, generated by spin-dependent scattering at skyrmions. Other factors which result in a decrease of the SkHA include damping value, with larger values resulting in lower SkHA both for SOT and ISTT+STT+SOT models, as may obtained from Equation (6) – this is also discussed in the Supplementary Information, and below.

In a recent work the SkHA was investigated in Pt(3 nm)/Co(1.2 nm)/MgO(1.5 nm) multilayered tracks, also as a function of skyrmion diameter, and using skyrmion collections [31]. The experimental results are remarkably similar to those shown here, also showing small SkHA values, nearly-independent of skyrmion diameter, and small velocities under 10 ms$^{-1}$. Previous works on multilayers incorporating HM/FM layers, e.g. Pt(3 nm)/Co(0.9 nm) [10], and Pt(3.2 nm)/CoFeB(0.7 nm) [6], have also shown small SkHA values at skyrmion diameters ~100 nm to 150 nm. Similarly in Ref. [7], low SkHA values were obtained (average ~15.2°) in Pt(2.7 nm)/CoFeB(0.86 nm)/MgO(1.5 nm) multilayers, with skyrmion diameters ~150 nm. These values could be reproduced by SOT-only modelling, only by assuming very large damping values, up to $\alpha = 0.5$. As shown in the Supplementary Information, this indeed results in the SOT-only SkHA values closer to those found in experiments, however a monotonic increase in SkHA with decreasing diameter still persists, with discrepancies remaining for sub-100nm diameter skyrmions. Moreover, a large body of work with measurements of damping in ultrathin Co, CoB, and CoFe films interfaced with Pt, reveal values of $\alpha \cong 0.1$ and lower [18,31,50-60], for typical FM layer thicknesses used for skyrmion studies. With an emerging consensus pointing to damping values $\alpha < 0.1$, large quantitative discrepancies would arise between experimental SkHA values [6,7,10], and SOT-only modelling, particularly for small skyrmion diameters. The large values of damping used in these works are derived from domain wall mobility measurements, for example as shown in Ref. [61], although some experimental evidence for large damping values was also obtained for Pt/Co/Pt trilayers in Ref. [62]. These damping values however stand in sharp contrast with those obtained from other measurement techniques. For example, in Ref. [31], FMR measurements have shown $\alpha = 0.05$, for Co(1.2 nm) on Pt(3 nm). Other FMR measurements obtained $\alpha$ in the range 0.02 to 0.1 for stacks containing Co with thickness in the range 1.1 nm down to 0.6 nm respectively, on Pt(3 nm) [60], $\alpha = 0.03$ for [Pt(1.5 nm)/Co(1 nm)/W(1.5 nm)]$_N$ [50], $\alpha = 0.02$ for [Co(0.5 nm)/Pt(0.3 nm)]$_{\times 6}$ [59], $\alpha < 0.1$ for [Co(1.4 nm)/Pt(1 nm)]$_8$ [51]. Spin-torque FMR measurements on stacks incorporating Co(1 nm)/Pt(4 nm) and CoFe(1 nm)/Pt(4 nm) have obtained $\alpha < 0.1$ [54]. BLS measurements on Pt(2 nm)/Co(0.6 nm – 3.6 nm)/Ir(2 nm), have shown $\alpha < 0.1$ [53]. Time-resolved magneto-optical Kerr effect measurements have shown $\alpha$ between 0.1 and 0.13 for [Co(0.4 nm)/Pt(0.8 nm)]$_N$ [52], whilst values in the range 0.03 to 0.06 were obtained for Pt(3 nm)/Co(0.8 nm)/HM trilayers (HM = W, Ta, Pd) [56]. In stacks containing Pt(2.7 nm)/CoB(0.8 nm)/Ir(0.4 nm), as in the current work, we've also previously determined $\alpha = 0.07$ by analysing larger diameter skyrmions [18]. A review of experimental results on damping in ferromagnetic thin films and multilayers is found in Ref. [63]. At large skyrmion diameters, as seen in Fig. 3 and the Supplementary Information, the SOT-only model is close to the full spin torque model in terms of SkHA, and SOT-only modelling can reproduce experimental values in this diameter range with realistic damping values. Thus in Ref. [9], skyrmion diameters were ~800 nm and larger, and experimental SkHA values were close to those obtained from SOT-only modelling with $\alpha = 0.02$, in Ta(5 nm)/CoFeB(1.1 nm).

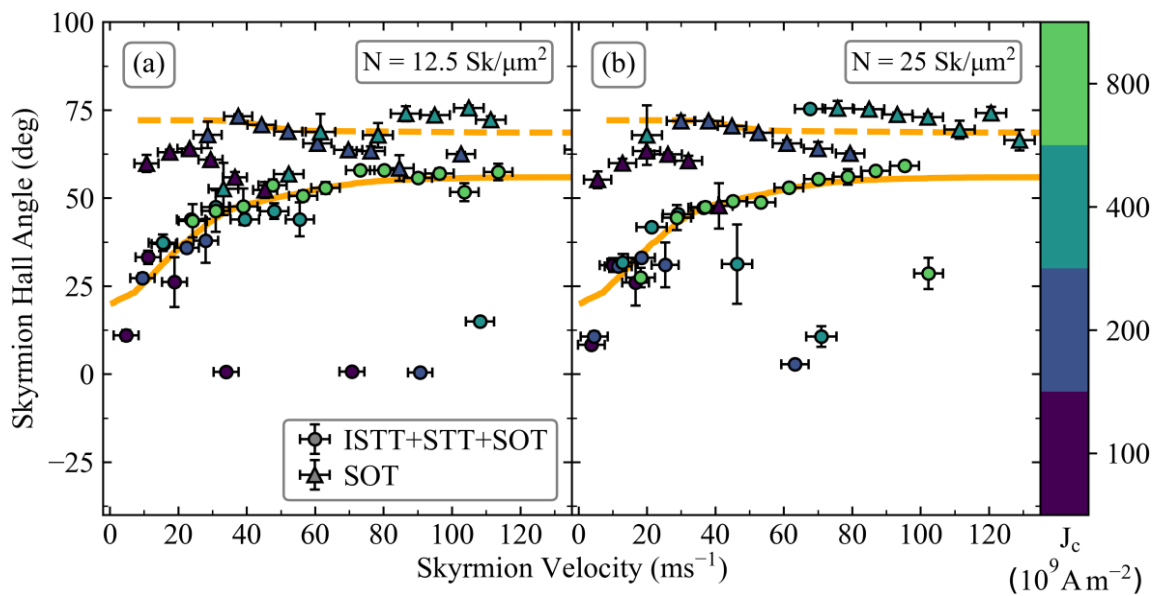

**Figure 4**. SkHA dependence on the skyrmion velocity with landscape disorder, for skyrmion collections with (a) $N = 12.5$ Sk/μm$^2$ and (b) $N = 25$ Sk/μm$^2$. The solid disks and triangles show results for ISTT+STT+SOT, and SOT-only respectively, for a corresponding applied current density indicated by the colour bar. The magenta diamonds indicate the mean value obtained from experiments (average $J_C$ = 2×10$^{11}$ A/m$^2$). The solid and dashed lines are Savitzky-Golay-fit trend lines, for ISTT+STT+SOT, and SOT respectively, acting as a guide to the eye.

It is well known the SkHA depends on the driving force when the effect of disorder is taken into account [7,9,10]. This is also reproduced for skyrmion collections, as shown in Fig.

4, both for the SOT-only results, and the full spin torque, although the effect is more pronounced for the latter. These results show an initial increase in the SkHA with increasing velocity, indicative of the depinning regime. At larger velocities the SkHA tends to a constant value, indicative of the flow regime, reaching the limiting SkHA values obtained using the Thiele model as expected. Whilst for the collective skyrmion motion a large spread of SkHA is obtained, as may be seen in Fig. 3, the results in Fig. 4 are velocity-bin-averaged. The spread in SkHA values is larger for the full spin torque model (in agreement with experimental results shown in Fig. 3(c)), since the combination of the various spin torques results in lower skyrmion velocities, and are therefore more strongly influenced by the local landscape disorder compared to the SOT-only case. In experiments containing over 4500 skyrmion paths, just over 3500 skyrmions are not pinned. From these, the average SkHA is ~24° ± 12°, with velocities ranging up to 2.5 ms$^{-1}$, and mean of 0.5 ms$^{-1}$. These values are in agreement with the trend obtained with ISTT+STT+SOT in Fig. 4, and is again in disagreement with SOT-only modelling. To obtain larger velocities in experiments larger current densities are required, however these degrade the samples over repeated measurements.

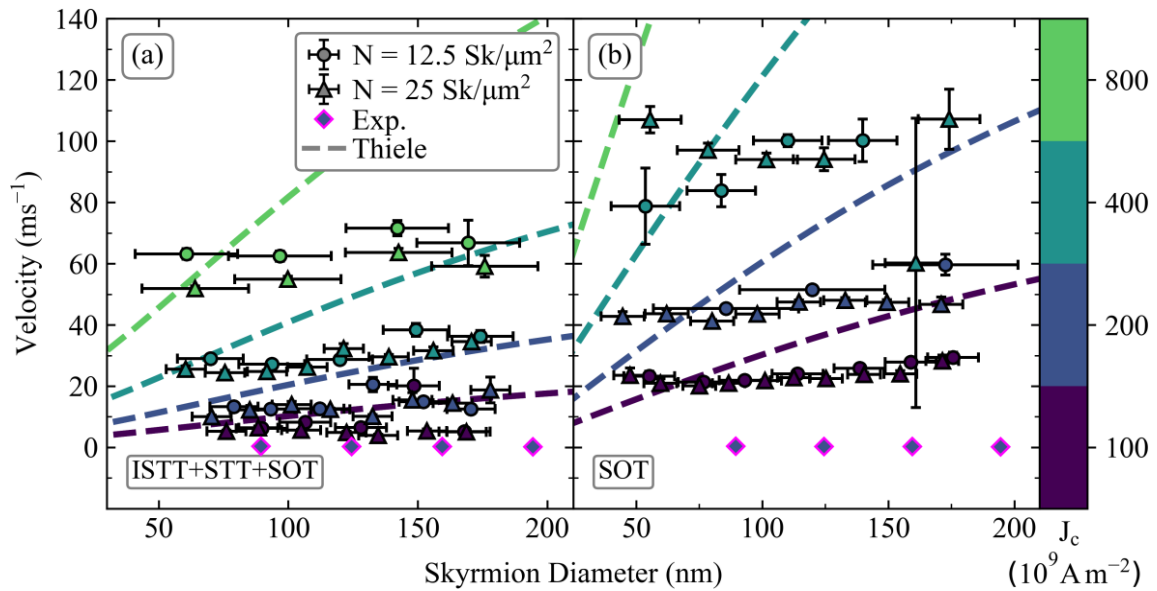

**Figure 5**. Skyrmion velocity diameter dependence. Panel (a) shows results for skyrmions under the influence of ISTT+STT+SOT and (b) shows the influence of SOT alone. The solid disks and triangles represent $N$ = 12.5, and $N$ = 25 Sk/µm$^2$ respectively, each corresponding to an applied current density shown by the colour bar. The magenta outlined diamonds represent the experimental results. The coloured dashed lines show the solutions obtained from the Thiele model with $\alpha$ = 0.1, for an isolated skyrmion at the respective current density, using the same color coding.

The skyrmion velocity-diameter dependence is shown in Fig. 5 for (a) ISTT+STT+SOT, and (b) SOT only. The symbols show results obtained for collective motion, whilst the dashed lines are obtained from the Thiele model (thus for isolated skyrmions without landscape disorder). For the isolated skyrmion motion the velocity increases monotonically with diameter as expected. For the skyrmion collection however, it is observed the velocity is nearly independent of diameter. This is to be expected, since the skyrmions move elastically, and due to repulsive interactions with neighbouring skyrmions an average group velocity is established. Thus, the larger diameter skyrmions, which in isolation move with larger velocities, are slowed down by smaller diameter skyrmions in the collection, and conversely the smaller diameter skyrmions move faster than they do in isolation, as they are pushed by the larger diameter skyrmions. This holds both for the SOT-only case, as well as the full spin torque case. Experimental skyrmion velocities recorded reach up to ~2.5 ms$^{-1}$ (mean ~0.5 ms$^{-1}$) which are lower than the values obtained in simulations. This is possibly due to the multilayer composition, interface quality, and multilayer stack repetition. However, with the inclusion of the ISTT and STT, the skyrmion velocities align closer to those found experimentally, compared to the SOT-only case, as may be seen in Fig. 5. Thus, for the full spin torque case at $J_C = 2\times10^{11}$ A/m$^2$, the average velocity obtained from modelling is ~10 ms$^{-1}$. On the other hand, for the SOT-only case the average velocity at this current density is > 40 ms$^{-1}$. We also note that in Ref. [31] average collective skyrmion velocities range up to ~10 ms$^{-1}$ (for $J_C$ up to ~5.5×10$^{11}$ A/m$^2$), which is again close to the values we obtained here using the full spin torque, but is in disagreement with the SOT-only model. Moreover, in Ref. [8] for stacks based on Pt(3 nm)/Co(0.6 nm – 1 nm)/Ir(1 nm), also containing skyrmion collections with sub-100 nm diameter skyrmions, velocities of less than 0.5 m/s were observed for $J_C$ up to 3.0×10$^{11}$ A/m$^2$. Once again, this is in disagreement with the SOT-only model, even when disorder is taken into account, but the values obtained here by including ISTT are closer to the experimental values. These results, put together with the SkHA values in Fig. 3(c), show the SOT-only model does not adequately account for the experimental data. Moreover, these results also highlight the importance of the ISTT, which helps to account for the discrepancy between the experimental results and SOT-only model, particularly for SkHA values, but also helps to explain the small experimental skyrmion velocity values.

Finally, we note that ISTT should also be present for domain walls in multilayers, since any magnetization gradients in FM layers will result in diffusive vertical spin currents towards adjacent metallic layers. We note that experimental evidence for a huge negative spin-transfer torque was obtained for ultrathin Co layers interfaced with Pt [64]. This torque was found to be much larger than the bulk STT, and act in the opposite direction (thus with a negative spin polarization), and scaling inversely with the Co layer thickness. These features are remarkably similar to those we discussed here in relation to ISTT for skyrmions, and we conjecture the origin of such a huge negative spin-transfer torque could be the ISTT arising from diffusive vertical spin currents. Investigation of ISTT for domain walls in multilayers, however, is left for a future work. Finally, we also note evidence for a STT with effective negative spin polarization was found in Ref. [65], for skyrmion bubble motion. As remarked by the authors, this is not compatible with SOT, but with a negative STT. Again, we conjecture the observed negative STT could be the interfacial contribution to the total STT, which indeed acts with an effective negative spin polarization. A possible further experiment to directly demonstrate the presence of ISTT, is to investigate a stack with broken inversion symmetry, e.g. Pt/Co/Cu/Pt, but with Pt layers of equal thickness. Such a structure may allow cancellation of the SOT from the Pt layers, due to the long spin-flip length in Cu, whilst maintaining the inversion asymmetry, and hence non-zero DMI, required to stabilize Néel skyrmions. In such a structure, any skyrmion movement cannot arise from SOT, allowing for direct comparison of skyrmion movement and ISTT modelling.

Conclusion

We have studied the motion of skyrmion collections in ultrathin HM/FM multilayers with landscape disorder, utilizing micromagnetic simulations, with spin torques computed with a self-consistent spin transport solver, and compared these to experimental results and numerical modelling using the Thiele equation. The SOT, ISTT, and STT have been considered, and we find that for ultrathin films the bulk STT effect on skyrmion motion is negligible compared to the SOT and ISTT. The SkHA is strongly dependent on the effect of the combined torques, and the small SkHA values observed in experiments are in good agreement with modelling which includes the ISTT, also explaining the observed diameter-independent SkHA, and spread of experimental SkHA values at the same current density. This is in contrast to the model which considers the SOT as the only driving torque, where large discrepancies in SkHA of 50° are observed. We also find the collective skyrmion motion is described by an average group velocity over the diameter range, which stands in marked contrast to the velocities obtained from isolated skyrmion motion. Furthermore, the group velocities obtained with the inclusion of ISTT are comparable to those found in experiments. This is not the case for the SOT-only model, which again shows large discrepancies of over 40 ms$^{-1}$ with the experimental data. Our results shed further light on collective current-induced skyrmion motion, and highlight the significance of the ISTT in understanding current-induced skyrmion motion in magnetic multilayers.

Data availability statement

The datasets generated during and/or analysed during the current study are available from the corresponding author on reasonable request.

Acknowledgements

Support from the European Union (H2020 grant MAGicSky No. FET-Open-665095.103 is gratefully acknowledged. Part of this work was carried out at the PolLux (X07DA) beamline of the Swiss Light Source, Paul Scherrer Institut (PSI), Switzerland. The PolLux end station was financed by the German Ministerium für Bildung und Forschung (BMBF) through contracts 05K16WED and 05K19WE2. This project has received funding from the EU-H2020 research and innovation programme under grant agreement No 654360 having benefitted from the access provided by having benefitted from the access provided by the Paul Scherrer Institut in Villigen within the framework of the NFFA-Europe Transnational Access Activity.

Author contributions

C.R.M. conceived the modelling methodology, performed the simulations, and analysed the data. K.Z. conceived the experimental methodology, grew the samples, performed the measurements and imaging, and analysed the data. S.F. and J.R. conceived the STXM imaging methodology and performed the imaging. C.H.M conceived and supervised the experiment. T.M. and P.R.B. analysed the modelling data and modelling methodology. S.L. conceived the model, performed part of the modelling, and analysed the data. S.L. and C.R.M. wrote the manuscript. All authors reviewed the manuscript.

Competing interests

The author(s) declare no competing interests.

Additional Information

Correspondence and requests for materials should be addressed to C.R.M and S.L.